\documentclass[aps,pra,twocolumn,floatfix,superscriptaddress]{revtex4-2}

\usepackage{amsmath}
\usepackage{amssymb}
\usepackage{graphicx}
\usepackage{dsfont}

\usepackage[unicode=true,
 bookmarks=true,bookmarksnumbered=false,bookmarksopen=false,
 breaklinks=false,pdfborder={0 0 1},backref=false,colorlinks=true]
 {hyperref}
\hypersetup{
 linkcolor=magenta, urlcolor=blue, citecolor=blue, pdfstartview={FitH}, unicode=true}

\usepackage{amsfonts}\usepackage{tabularx}\usepackage{dcolumn}\usepackage{bm}\usepackage{graphicx}
\usepackage{epstopdf}
\usepackage{xcolor}
\hypersetup{urlcolor=blue}
\usepackage{times}

\usepackage{float}
\makeatletter
\let\newfloat\newfloat@ltx
\makeatother

\usepackage{algorithm}
\usepackage{algorithmic}

\usepackage{placeins}

\newcommand{\cb}{\textcolor{blue}}

\begin{document}

\title{Efficient Quantum State Tomography with Mode-assisted Training}
\author{Yuan-Hang Zhang}
\email{yuz092@ucsd.edu}
\affiliation{Department of Physics, University of California, San Diego, CA 92093, USA}
\author{Massimiliano Di Ventra}
\email{diventra@physics.ucsd.edu}
\affiliation{Department of Physics, University of California, San Diego, CA 92093, USA}

\begin{abstract}

Neural networks (NNs) representing quantum states are typically trained using Markov chain Monte Carlo based methods. However, unless specifically designed, such samplers only consist of local moves, making the slow-mixing problem prominent even for extremely simple quantum states. 
Here, we propose to use mode-assisted training that provides global information via the {\it modes} of the NN distribution. Applied to quantum state tomography using restricted Boltzmann machines, this method improves the quality of reconstructed quantum states by orders of magnitude. The method is applicable to other types of NNs and may efficiently tackle problems previously unmanageable.

\end{abstract}
\maketitle

\section{Introduction}

With the ability to compress and extract information from high-dimensional data, machine learning has become a useful tool in a wide variety of fields~\cite{goodfellow2016deep}. Physics is no exception. 

For instance, neural networks (NNs) have been used with reasonable success as variational 
wavefunctions of quantum many-body systems \cite{carleo2017solving, torlai2018latent, gao2017efficient, torlai2018neural, cai2018approximating, carrasquilla2019reconstructing, cha2021attention, schmale2021scalable, ahmed2021quantum}. Irrespective of the type of NN employed as 
variational state, the vast majority of methods to train NNs rely on Markov chain Monte Carlo (MCMC) sampling~\cite{goodfellow2016deep}, which, unless specifically designed, only consists of local moves. As a result, the slow-mixing problem arises, significantly slowing down the algorithm, sometimes causing the training to fail completely, even for very simple systems. Countless efforts have been devoted to solving this problem, and various improved MCMC routines have been proposed, aiming at accelerating the mixing of the Markov chain \cite{tieleman2008training, earl2005parallel, desjardins2010parallel, cho2010parallel, sminchisescu2003mode, guan2007small}. Yet, they all serve one purpose: to increase the quality of the MCMC samples for a more accurate gradient estimation.

In fact, the cost function of an NN defines a non-convex landscape, and as any non-convex landscape, convergence to the global minimum with gradient-based methods can never be guaranteed. In some cases this may not be of concern, since proper design of the NN, weight initialization and/or learning rate scheduling empirically seem to guarantee a smooth convergence to some local minimum, close enough to the global one. However, as we will show below, conventional sampling methods can easily lead to bad local minima for certain types of quantum states with strongly non-local features, which could become a serious problem, causing complete failure of the training. 

In this paper, we tackle this issue from a new perspective: we design an {\it off-gradient} training step (i.e., a training step that does not align with the direction of the gradient), constructed using the {\it mode} of the NN distribution, which we call mode-assisted training \cite{manukian2020mode, manukian2021mode}. This method supplements the regular gradient descent with mode updates, which explicitly inject global information to the training process, leading to better estimations of the global minimum.


As an example of NNs, we will employ the well-known restricted Boltzmann machines (RBMs), and focus on the challenging task of reconstructing a quantum state with repeated measurements on its identical copies. This is called quantum state tomography (QST) \cite{fano1957description, hradil1997quantum}. While traditional, brute-force methods require tens of thousands of measurements to reconstruct even small quantum states  \cite{haffner2005scalable}, recent advancements in machine  learning methods have greatly improved the efficiency of such a task, making it feasible to perform QST on states with tens or even hundreds of qubits~\cite{torlai2018neural, carrasquilla2019reconstructing, quek2021adaptive, schmale2021scalable}. Yet, as we will show below, such methods are still inefficient when the quantum states showcase strongly non-local features. Instead, mode-assisted training significantly improves the quality of reconstructed quantum states while reducing the number of required measurements by orders of magnitude. This opens up the possibility of efficiently tackling other types of quantum problems previously unmanageable with these types of approaches. 

\section{Restricted Boltzmann machines}
In this section, we outline the basics of the RBM, and leave the detailed calculations in Appendix~\ref{sec:RBM_intro}. As illustrated in Fig.~\ref{fig:RBM}, an RBM is a two-layered neural network with $n$ visible nodes $\mathbf{v}\in\{0, 1\}^n$, $m$ hidden nodes $\mathbf{h}\in\{0, 1\}^m$, and trainable weights $W_{ij}$ and biases $a_i$, $b_j$. Together, they define a joint distribution, 

\begin{equation}
    p(\mathbf{v}, \mathbf{h})=\frac{1}{Z} \exp\left(\sum_{i=1}^n a_i v_i + \sum_{j=1}^m b_j h_j + \sum_{i=1}^n \sum_{j=1}^m W_{ij} v_i h_j\right), \label{eq:RBM_joint}
\end{equation}
where $Z$ is the partition function. The marginal distribution of the visible nodes, 
\begin{equation}
		p(\mathbf{v}) = \sum_{\mathbf{h}} p(\mathbf{v}, \mathbf{h})
		=\frac{1}{Z}e^{\sum_i a_i v_i} \prod_{j}\left(1+e^{b_j+\sum_i v_i W_{ij}}\right),\label{eq:RBM}
\end{equation}
%
is used to model the unknown data distribution.

\begin{figure}[htbp]
    \centering
    \includegraphics[width = 0.25\textwidth]{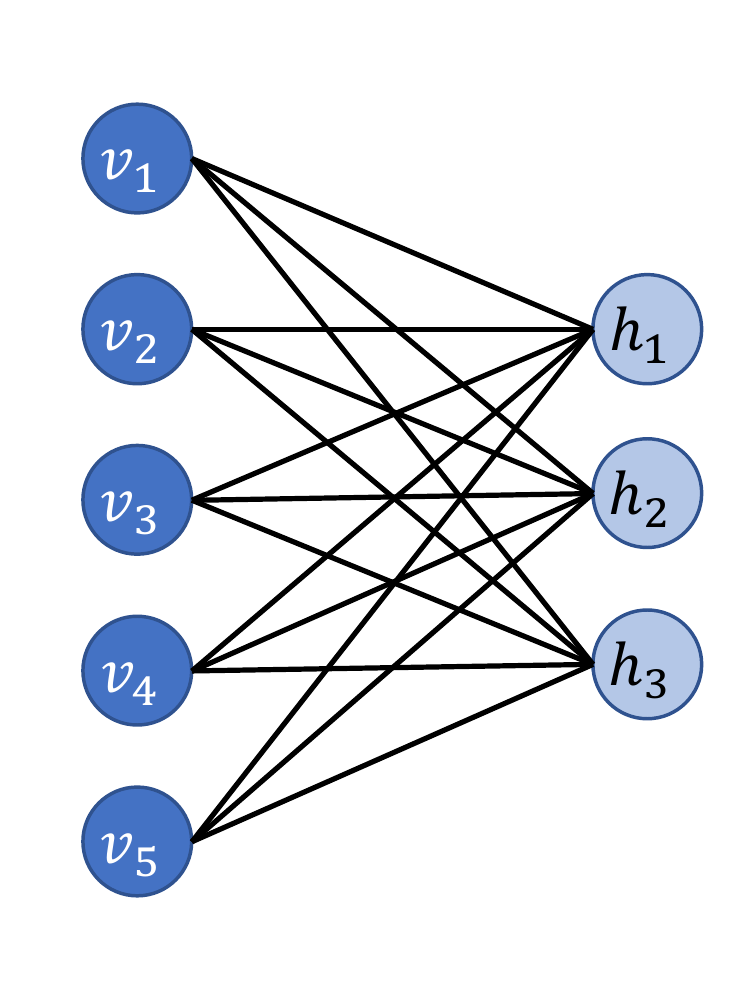}
    \caption{The structure of a typical RBM. The visible nodes $v_i$ and hidden nodes $h_j$ form a bipartite graph, with no intra-layer connections.}
    \label{fig:RBM}
\end{figure}

To represent a quantum state, $|\psi\rangle$, we map the wavefunction to such a probability distribution. For instance, if the wave function is positive, we can simply set $\psi(\mathbf{v})=\sqrt{p(\mathbf{v})}$. For a general complex wavefunction, one can either allow the RBM weights to have complex values \cite{carleo2017solving}, or use a second RBM to model the phase \cite{torlai2018latent}. Yet another approach is to model the quantum state with informationally complete positive operator-valued measurements (IC-POVM) \cite{carrasquilla2019reconstructing}, whose outcome is an ordinary probability distribution instead of a quasi-probability distribution. 

Irrespective, the standard method of training an RBM is to minimize the KL divergence \cite{hinton2002training, fischer2012introduction}, 
\begin{equation}
    \mathrm{KL}(q||p) = \sum_{\mathbf{v}}q(\mathbf{v})\log\frac{q(\mathbf{v})}{p(\mathbf{v})},
\end{equation}
between the target distribution $q(\mathbf{v})$ and the RBM distribution $p(\mathbf{v})$. Computing the gradient with respect to the RBM parameters, we obtain the formula of weight updates: 

\begin{equation}
    \Delta W_{ij} = \eta \Big(\langle v_i h_j\rangle_{q(\mathbf{v})p(\mathbf{h}|\mathbf{v})} - \langle v_i h_j\rangle_{p(\mathbf{v}, \mathbf{h})}\Big),
    \label{eq:gradient}
\end{equation}
in which $\eta$ is the learning rate, and $\langle \cdot \rangle_{p}$ denotes expectation over the probability distribution $p$. A similar expression holds for the biases~\footnote{Throughout this work, $\eta=0.01$ and reduces by half whenever performance doesn't improve for $10^4$ iterations, and the minibatch size for computing the expectation is $N^2$, where $N$ is the number of qubits.}. See Appendix~\ref{sec:RBM_intro} for detailed derivations.

In Eq.~\eqref{eq:RBM_joint}, the partition function $Z$ involves a summation over an exponential amount of terms, making it impossible to evaluate $p(\mathbf{v}, \mathbf{h})$ efficiently. Instead, $Z$ cancels out in the conditional probabilities, $p(\mathbf{h}|\mathbf{v})$ and $p(\mathbf{v}|\mathbf{h})$, making them efficiently computable \cite{fischer2012introduction}:
\begin{equation}
\begin{aligned}
    p(\mathbf{h}|\mathbf{v})=&\prod_{j=1}^m \frac{e^{h_j(b_j+\sum_{i=1}^n v_i W_{ij})}}{1+e^{b_j+\sum_{i=1}^n v_i W_{ij}}}\\
    p(\mathbf{v}|\mathbf{h})=&\prod_{i=1}^n \frac{e^{v_i(a_i+\sum_{j=1}^m W_{ij}h_j)}}{1+e^{a_i+\sum_{j=1}^m W_{ij}h_j}}
\end{aligned}
\end{equation}
Therefore, in the expression of the gradient, Eq.~\eqref{eq:gradient}, the first expectation can be evaluated exactly and efficiently, while the second expectation is usually approximated with a sampling algorithm.

\section{Local Samplers}

Contrastive divergence (CD)~\cite{goodfellow2016deep} is the most widely adopted sampling algorithm for RBMs. CD-$k$ starts from a sample $\mathbf{v}^0$ from the dataset and constructs a Markov chain of samples,
\begin{equation}
    \mathbf{v}^0\to\mathbf{h}^0\to\mathbf{v}^1\to\mathbf{h}^1\to\cdots\to\mathbf{v}^k, \label{eq:chain}
\end{equation}
using the conditional distributions $p(\mathbf{h}|\mathbf{v})$ and $p(\mathbf{v}|\mathbf{h})$, alternating between the visible nodes $\mathbf{v}$ and hidden nodes $\mathbf{h}$ for $k$ times. When $k\to\infty$, the distribution of $\mathbf{v}^k$ converges to $p(\mathbf{v})$, and we can approximate the second term in Eq.~\eqref{eq:gradient} with an expectation over a batch of sampled $\mathbf{v}^k$.

CD, among many other Markov chain based samplers, like persistent contrastive divergence (PCD) \cite{tieleman2008training} and parallel tempering (PT) \cite{earl2005parallel, desjardins2010parallel, cho2010parallel}, belongs to the category of ``local samplers''. Unless specifically designed, they only contains ``local moves'' and does not include global information on the probability distribution. In Appendix~\ref{sec:distance}, we rigorously define the concept of locality with respect to Markov chains, and visualize the spatial proximity of basis states over an example RBM. Here, we first demonstrate the potential issues that can arise with local samplers.

As a simple example, let us consider the Greenberger-Horne-Zeilinger (GHZ) state) \cite{greenberger1989going}, 
\begin{equation}
	|\Psi_\mathrm{GHZ}\rangle = \frac{1}{\sqrt{2}}\left(|00\cdots0\rangle + |11\cdots1\rangle\right),
\end{equation}
a prototypical $N$-qubit entangled state with two modes that has wide applications in quantum information theory, and is also used for benchmarking different QST algorithms \cite{carrasquilla2019reconstructing, cha2021attention}. 

\begin{figure}[htbp]
    \centering
    \includegraphics[width = 0.48\textwidth]{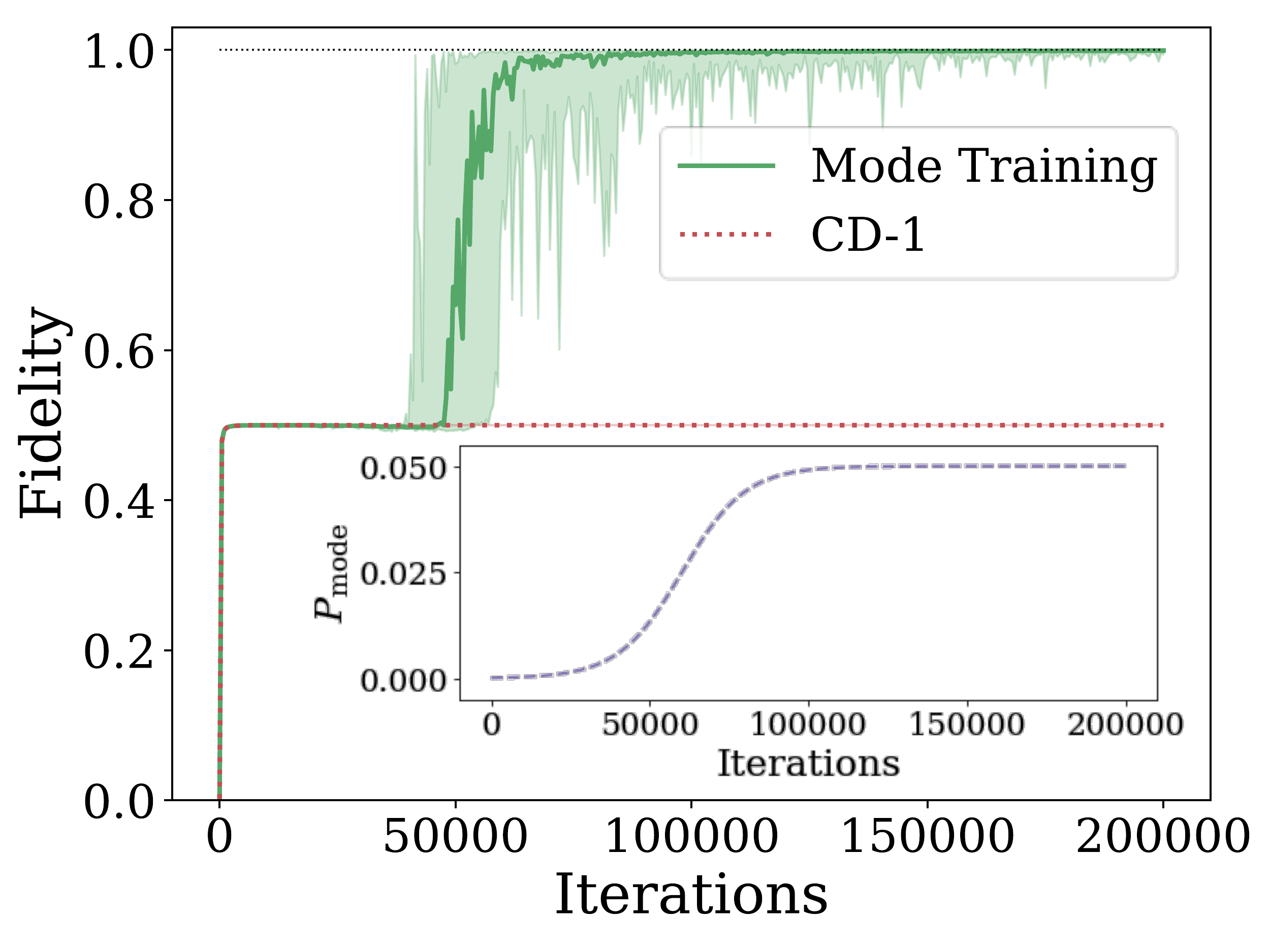}
    \caption{Training an RBM to represent a 10-qubit GHZ state. Training data is obtained by performing projective measurements in the $\{|0\rangle, |1\rangle\}$ basis on $10^4$ copies of the state. The inset shows the frequency of mode updates, Eq.~(\ref{eq:p_mode}) (with parameters $\alpha=20$, $\beta=6$ and $P_{\text{max}}=0.05$), as training goes on. Curves represent the medians of 20 runs, and the shaded regions are enclosed by the maximum and minimum. }
    \label{fig:GHZ}
\end{figure}

Superficially, this state seems trivial and has an exact RBM representation with only one hidden neuron~\cite{torlai2018neural}. On the other hand, training an RBM with CD to represent this state faces immediate failure. Starting from an ideal 10-qubit GHZ state, we obtain training data by performing projective measurements in the $\{|0\rangle, |1\rangle\}$ basis on $10^4$ copies of the state.  Fig.~\ref{fig:GHZ} (red dotted curve) shows the reconstruction fidelity for this state between the exact state $|\psi_\mathrm{exact}\rangle$ and the reconstructed one $|\psi\rangle$, $f(\psi_\mathrm{exact}, \psi)=|\langle \psi_\mathrm{exact} |\psi\rangle|^2$, for 10 qubits. CD can only learn one of the two modes, reaching a final fidelity of 1/2. As already mentioned, this is of no surprise, since CD is a local sampling algorithm and has difficulty mixing between different modes~\cite{sminchisescu2003mode}. If a CD chain starts from one mode, it is (almost) trapped there forever. This sampling bias is amplified over time, causing the RBM to eventually converge towards one of the two modes. 

To solve this problem, we need to properly incorporate global information into the training process. One simple yet efficient approach is mode-assisted training \cite{manukian2020mode}.

\section{Mode-assisted training}
\subsection{Algorithm}
To explicitly inject global information into the training process, we design an off-gradient training step that supplements the regular gradient updates, using the mode of the RBM distribution \cite{manukian2020mode}.
This means replacing the formula of weight updates, Eq.~(\ref{eq:gradient}), as follows:

\begin{equation}
    \Delta W^{\textrm{mode}}_{ij} = \eta \Big(\langle v_i h_j\rangle_{q_{\mathrm{mode}}(\mathbf{v})p(\mathbf{h}|\mathbf{v})} - [v_i h_j ]_{p(\cb{\mathbf{v},\mathbf{h}})}\Big)
    \label{eq:mode}
\end{equation}
where $[\cdot]$ indicates expectation over the mode of the RBM distribution $p(\cb{\mathbf{v},\mathbf{h}})$, and $q_{\mathrm{mode}}$ is a uniform distribution over all possible modes of the data distribution $q$. We call Eq.~\eqref{eq:mode} the ``mode-assisted training", or ``mode training" for short.

We train the NN for $n_{\max}$ iterations ($n_{\max}=2\times10^5$ throughout this work), and the schedule of when to perform mode updates is determined by calculating the probability of replacing Eq.~\eqref{eq:gradient} with Eq.~\eqref{eq:mode} at each training iteration step $n$ as: 
\begin{equation}
	\label{eq:p_mode}
	P_{\text{mode}}(n) = P_{\text{max}}\sigma\left(\alpha \frac{n}{n_{\max}} - \beta\right),
\end{equation}
where $\sigma$ is the sigmoid function, and $0 < P_{\text{max}}\leq 1$ is the maximum probability of a mode update. At the beginning of the training, $P_\text{mode}$ is then small, but it increases gradually with the number of updates, according to the parameters $\alpha$ and $\beta$ (see inset of Fig.~\ref{fig:GHZ})~\footnote{The choice of the sigmoid in 
Eq.~(\ref{eq:p_mode}) is arbitrary: other types of functions can be chosen to accomplish the same task.}.


To find $\mathbf{v}_{\mathrm{mode}}$, the mode of the RBM distribution, for distributions with a simple data structure, it suffices to evaluate the amplitudes of mode candidates from the dataset (e.g., $|00\cdots 0\rangle$ and $|11\cdots 1\rangle$ for the GHZ state). For more complicated distributions where identifying the mode candidates is hard, $q_{\mathrm{mode}}(\mathbf{v})$ can be approximated with $q(\mathbf{v})$, and $\mathbf{v}_{\mathrm{mode}}$ can be sampled from the joint distribution $p(\mathbf{v}, \mathbf{h})$, by employing an optimization solver. This involves minimizing the RBM energy, 
\begin{equation}
    E(\mathbf{v}, \mathbf{h})=-\sum_i a_iv_i -\sum_j b_jh_j-\sum_{i, j}W_{ij}v_i h_j,
\end{equation}
which is a quadratic unconstrained binary optimization problem \cite{kochenberger2014unconstrained} and is generally NP-hard. We employ the MemComputing solver \cite{traversa2018evidence, manukian2019accelerating, manukian2020mode,MemBook}, which can efficiently generate good approximations of $\mathbf{v}_{\mathrm{mode}}$, and an empirical polynomial scaling for typical RBMs is observed \cite{manukian2020mode}. 

We now show that when properly combined with regular CD updates, this off-gradient mode update greatly increases the stability of the training. This is clearly seen in Fig.~\ref{fig:GHZ}. In the initial phase of the training, CD is able to learn the support of the distribution but not much else. As training goes on, the frequency of mode updates is increased to balance between the multiple modes and bring the RBM distribution as close to the data distribution as possible. In fact, as the probability of mode updates increases, mode-assisted training easily jumps out of the local minimum and learns the full distribution with near perfect fidelity. 

\subsection{W-state}

\begin{figure}[htbp]
	\centering
	\includegraphics[width = 0.43\textwidth]{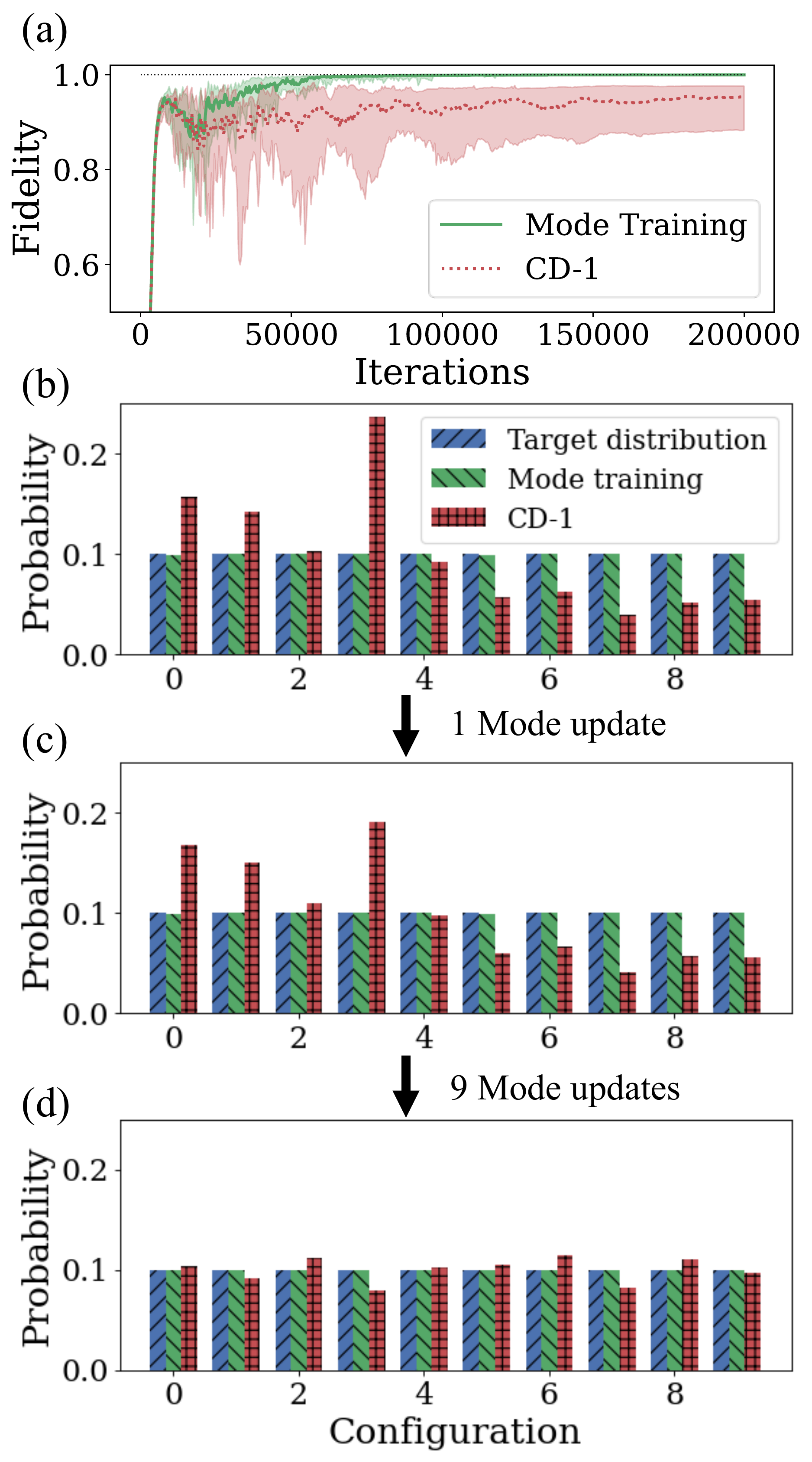}
	\caption{Training an RBM on the 10-qubit W state. Training data is obtained by performing projective measurements in the $\{|0\rangle, |1\rangle\}$ basis on $10^4$ copies of the state. (a) The fidelity curve during training. Curves represent medians of 20 runs, and the shaded regions are enclosed by the maximum and minimum. Mode-assisted training converges quickly to $f=1$ with vanishing variance, leaving a significant gap compared to CD-1. (b) Example of the distribution learned by the RBM, after training completes. Among all $2^{10}$ possible configurations, we only plotted the 10 most important ones in the W-state, $|10\cdots0\rangle$ through $|00\cdots1\rangle$. While the result of mode-assisted training is almost perfect, CD-1 produces far less satisfying results. (c) and (d) Smoothing the noisy distribution learned with CD-1 using additional mode updates. Each mode update locates the global maximum and ``pushes it'' down, while all other states ``pop up'' a bit. Repeated mode updates would eventually enforce uniformity over multiple modes.}
	\label{fig:W_mode_demo}
\end{figure}

In order to show even more clearly that mode-assisted training provides global (long-range) information during training, we consider the W-state (named after Wolfgang D\"{u}r) \cite{dur2000three}, another $N$-qubit entangled state given by
\begin{equation}
    |\Psi_W\rangle = \frac{1}{\sqrt{N}}\left(|100\cdots0\rangle + |010\cdots0\rangle + \cdots + |000\cdots1\rangle\right).
\end{equation}

Similar to the GHZ state, we constructed a synthetic dataset by taking $10^4$ projective measurements in the $\{|0\rangle, |1\rangle\}$ basis on a 10-qubit W-state, and trained two RBMs using CD-1 and mode-assisted training, respectively. Fig.~\ref{fig:W_mode_demo}(a) compares their reconstruction fidelity. In Fig.~\ref{fig:W_mode_demo}(b), we show the amplitude of the state after training: mode-assisted training quickly converges to the target distribution almost perfectly, but the result for CD-1 is a lot more noisy. Again, while CD-1 can correctly locate the mode states (the support of the distribution), it does a terrible job at matching their amplitudes to the target. We included two GIF files in the supplemental materials, visualizing the entire training process from scratch, from which the advantage of mode-assisted training is clearly visible.

At this point, if we correct the noisy distribution learned with CD-1, we can visualize how mode-assisted training works. Figs.~\ref{fig:W_mode_demo}(b), (c) and (d) show this procedure, where we applied 10 mode updates to the distribution learned with CD-1. The correction is already 
significant after a single mode update: from (b) to (c), the global maximum of the CD-1 distribution is pushed down, while all other states pop up a bit. Repeating this procedure for an additional 9 times, the resulting distribution is already very close to uniformity.  The direct access to global maxima is what local samplers like CD lack, and this deficiency cannot be solved by increasing the length of the sampling chain. This is explicitly shown Appendix~\ref{sec:slow_mix}. In Appendix~\ref{sec:PCD_PT}, we compare the performance of mode-assisted training against persistent contrastive divergence \cite{tieleman2008training} and parallel tempering \cite{earl2005parallel, desjardins2010parallel, cho2010parallel}, two advanced (albeit still local) samplers that are designed to alleviate the slow-mixing problem. Several additional numerical experiments are presented in Appendix~\ref{sec:numerical_results}, showing the performance of mode training under different scenarios. 

\subsection{Scalability}

\begin{figure}[htbp]
	\centering
	\includegraphics[width = 0.45\textwidth]{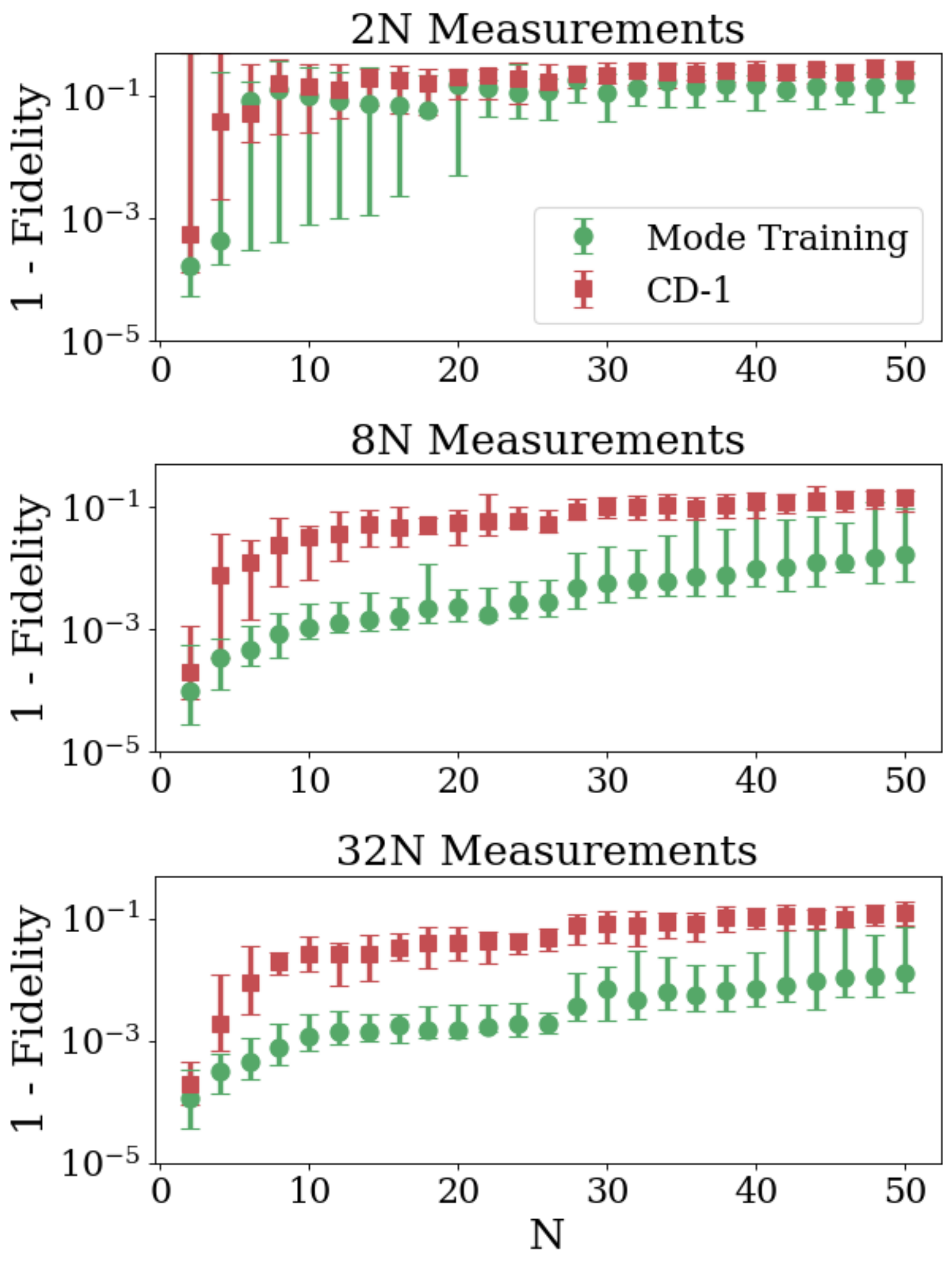}
	\caption{Comparing the reconstruction fidelity of CD-1 and mode-assisted training for the W-state with $N$ qubits and fixed amount of measurements. Data points are computed using the median of 20 runs, and error bars represent corresponding maximum and minimum values. In all cases, mode-assisted training outperforms CD-1, and the difference is increasing as we increase the number of measurements, up to two orders of magnitude.}
	\label{fig:f_fixed_measure}
\end{figure}

So far, we have only considered small systems. Now, we scale up the system size and show that mode training requires orders of magnitude less number of measurements compared to MCMC sampling. As an explicit example we consider the W-state up to 50 qubits, and focus on two physical quantities: fidelity and number of measurements. To study the effect of measurements on reconstruction quality, we mimic experiments by first performing a fixed number of measurements on the exact state, then use the measured results as dataset to train the RBM model. As shown in Fig.~\ref{fig:f_fixed_measure} for a W-state with $N$ qubits, when fixing the number of measurements, mode training outperforms CD-1 by one to two orders of magnitude. 

\begin{figure}[htbp]
	\centering
	\includegraphics[width = 0.45\textwidth]{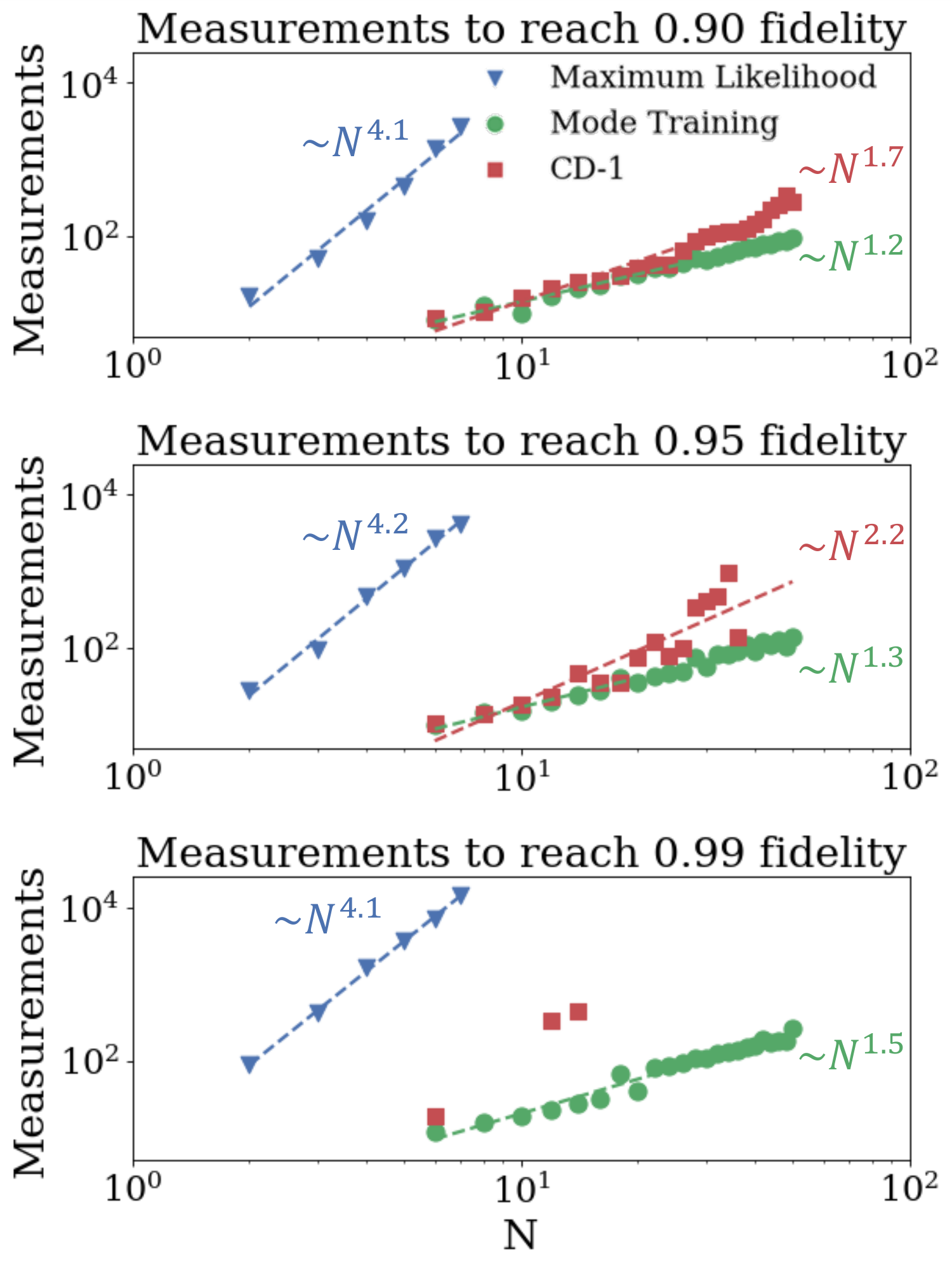}
	\caption{Number of measurements required to reach a certain fidelity for the W state with $N$ qubits with maximum likelihood, CD-1, and mode-assisted training. Data points are computed with interpolation, using the best result from 20 runs with fixed number of measurements.  }
	\label{fig:measure_to_f}
\end{figure}

Next, we fix a target fidelity and estimate the amount of measurements required to reach that fidelity. For better comparison, we also included results from the maximum likelihood method, a brute-force approach for quantum state tomography \cite{hradil1997quantum}. The difference is drastic: maximum likelihood has a hard time reaching any fidelity using reasonable resources already for less than 10 qubits. This is due to the exponential space complexity of storing and manipulating the full density matrix, which allows us to show results only up to 7 qubits in Fig.~\ref{fig:measure_to_f}. 

When the fidelity target is not too high ($f\lesssim 0.95$), CD-1 shows a performance comparable to mode-assisted training. However, as the target fidelity increases, CD increasingly struggles to reach the same fidelity, eventually disappearing from the plot due to its inability to reach the target with less than $10^4$ measurements. Mode-assisted training, on the other hand, performs consistently throughout the size range, and shows a sub-quadratic scaling (see Fig.~\ref{fig:measure_to_f}) in the number of measurements to reach the target up to the size considered, without much sensitivity to the target fidelity. 

\section{Conclusion}
In this work, we have shown that providing global information to the training of an NN representing quantum states, in the form of the modes of its probability distribution, improves significantly the reconstruction of such states. The improvement also translates into orders of magnitude reduction in the number of required measurements. We have employed RBMs as example, but the method is applicable to other types of NNs \cite{manukian2021mode}. 

We have also shown that the mode-assisted training method scales very favorably in terms of number of measurements required to reach a target fidelity as a function of number of qubits. This result, coupled with optimization methods, like MemComputing~\cite{MemBook}, to efficiently sample the mode(s) of multi-dimensional probability distributions, paves the way to solve a wide variety of quantum problems 
classically and with considerably less resources.

\emph{Data availability}\textemdash 
The code for all simulations performed in this paper is available at \href{https://github.com/yuanhangzhang98/mode_qst}{https://github.com/yuanhangzhang98/mode\_qst}. 

\emph{Acknowledgments}\textemdash
We acknowledge financial support from the Department of Energy under Grant No. DE-SC0020892.

\bibliographystyle{apsrev4-2}
\bibliography{SUSYref}

\clearpage
\newpage
\appendix


\section{An introduction to restricted Boltzmann machines}
\label{sec:RBM_intro}

For completeness, in this section, we systematically introduce the restricted Boltzmann machine (RBM)~\cite{goodfellow2016deep}. 

Fig.~\ref{fig:RBM} shows the structure of a typical RBM, where two sets of binary nodes, $\{v_i, h_j\}$, lies on a bipartite graph. One can view the RBM as a classical Ising model, with each binary node as an individual spin. The weight matrix $W_{ij}$ parameterizes the interaction between the spins, and each spin has an external field, $a_i$ or $b_j$, acting on it. Combining everything, we can define the energy of the RBM: 

\begin{equation}
    E(\mathbf{v}, \mathbf{h})=-\left(\sum_{i=1}^n a_i v_i + \sum_{j=1}^m b_j h_j + \sum_{i=1}^n \sum_{j=1}^m W_{ij} v_i h_j\right)
\end{equation}

At equilibrium, the distribution of the spins is characterized by the Boltzmann distribution (hence the name Boltzmann machine): 

\begin{equation}
    p(\mathbf{v}, \mathbf{h})=\frac{1}{Z}e^{-E(\mathbf{v}, \mathbf{h})}
\end{equation}
where $Z=\sum_{\mathbf{v}, \mathbf{h}}e^{-E(\mathbf{v}, \mathbf{h})}$ is the partition function. To model an unknown distribution, we use the marginal distribution of $\mathbf{v}$, 
\begin{equation}
    p(\mathbf{v}) = \sum_{\mathbf{h}} p(\mathbf{v}, \mathbf{h}).
    \label{eq:sum}
\end{equation}

There are two conventions when choosing the values of the binary nodes, either $\{0, 1\}$ or $\{+1, -1\}$, and conversion between the two representations can be easily carried out via a transformation on the weights and biases \cite{pei2020generating}. We stick to the former convention, $\{\mathbf{v}, \mathbf{h}\}\in\{0, 1\}^{n+m}$. Then, the summation in Eq.~\eqref{eq:sum} can be carried out explicitly: 
\begin{equation}
\begin{aligned}
    p(\mathbf{v}) =& \frac{1}{Z}\sum_{\mathbf{h}\in\{0,1\}^m} e^{-E(\mathbf{v}, \mathbf{h})}\\
    =&\frac{1}{Z}e^{\sum_{i=1}^n a_i v_i} \prod_{j=1}^m\left(1+e^{b_j+\sum_{i=1}^n v_i W_{ij}}\right)
\end{aligned}\label{eq:pv}
\end{equation}

The partition function $Z$ involves a summation over an exponential amount of terms, making it impossible to evaluate Eq.~\eqref{eq:pv} exactly. Instead, the conditional probabilities, $p(\mathbf{h}|\mathbf{v})$ and $p(\mathbf{v}|\mathbf{h})$, can be computed efficiently: 

\begin{equation}
\begin{aligned}
    p(\mathbf{h}|\mathbf{v})=&\frac{p(\mathbf{v}, \mathbf{h})}{p(\mathbf{v})}\\
    =& \prod_{j=1}^m \frac{e^{h_j(b_j+\sum_{i=1}^n v_i W_{ij})}}{1+e^{b_j+\sum_{i=1}^n v_i W_{ij}}}\\
    =& \prod_{j=1}^m p(h_j|\mathbf{v})
\end{aligned}
\end{equation}
with
\begin{equation}
\begin{aligned}
    &p(h_j=1|\mathbf{v})=\mathrm{sigmoid}(b_j+\sum_{i=1}^n v_i W_{ij}),\\
    &p(h_j=0|\mathbf{v})=\mathrm{sigmoid}(-b_j-\sum_{i=1}^n v_i W_{ij}).
\end{aligned} \label{eq:phv}
\end{equation}

Thanks to the bipartite structure of RBM, with fixed $\mathbf{v}$, different $h_j$ are independent of each other, and the conditional probability $p(\mathbf{h}|\mathbf{v})$ factors into a product form. 
Similarly, 

\begin{equation}
\begin{aligned}
    p(\mathbf{v}|\mathbf{h})&=\prod_{i=1}^n p(v_i|\mathbf{h}),\\
    p(v_i=1|\mathbf{h})&=\mathrm{sigmoid}(a_i+\sum_{j=1}^m W_{ij}h_j),\\
    p(v_i=0|\mathbf{h})&=\mathrm{sigmoid}(-a_i-\sum_{j=1}^m W_{ij}h_j). 
\end{aligned}\label{eq:pvh}
\end{equation}

Eqs.~\eqref{eq:phv}, \eqref{eq:pvh} can be efficiently computed and are frequently used in the training and sampling procedures. 

To model the target distribution $q(\mathbf{v})$, we train the RBM by minimizing the KL divergence \cite{hinton2002training, fischer2012introduction}, 
\begin{equation}
    \mathrm{KL}(q||p) = \sum_{\mathbf{v}}q(\mathbf{v})\log\frac{q(\mathbf{v})}{p(\mathbf{v})}.     
\end{equation}

Explicitly computing the gradients with respect to the RBM weights, we have: 
\begin{equation}
\begin{aligned}
    &\frac{\partial \mathrm{KL}(q||p)}{\partial W_{ij}}\\
    =& -\sum_{\mathbf{v}}q(\mathbf{v})\frac{1}{p(\mathbf{v})}\frac{\partial p(\mathbf{v})}{\partial W_{ij}}\\
    =& -\sum_{\mathbf{v}}q(\mathbf{v}) \frac{1}{p(\mathbf{v})}\sum_{\mathbf{h}}\frac{\partial p(\mathbf{v}, \mathbf{h})}{\partial W_{ij}}\\
    =& -\sum_{\mathbf{v}, \mathbf{h}}q(\mathbf{v})\frac{p(\mathbf{v}, \mathbf{h})}{p(\mathbf{v})}v_i h_j + \frac{1}{Z}\frac{\partial Z}{\partial W_{ij}}\\
    =& -\sum_{\mathbf{v}, \mathbf{h}}q(\mathbf{v})p(\mathbf{h}|\mathbf{v}) v_i h_j + \sum_{\mathbf{v}, \mathbf{h}}p(\mathbf{v}, \mathbf{h}) v_i h_j\\
    =& -\langle v_i h_j\rangle_{q(\mathbf{v})p(\mathbf{h}|\mathbf{v})} + \langle v_i h_j\rangle_{p(\mathbf{v},\mathbf{h})}\label{eq:gradient_full}
\end{aligned}
\end{equation}

Similarly, 
\begin{equation}
\begin{aligned}
    &\frac{\partial \mathrm{KL}(q||p)}{\partial a_i} = -\langle v_i \rangle_{q(\mathbf{v})} + \langle v_i \rangle_{p(\mathbf{v},\mathbf{h})}\\
    &\frac{\partial \mathrm{KL}(q||p)}{\partial b_j} = -\langle  h_j\rangle_{q(\mathbf{v})p(\mathbf{h}|\mathbf{v})} + \langle  h_j\rangle_{p(\mathbf{v},\mathbf{h})}    
\end{aligned}
\end{equation}

With the analytic expression of the gradients, we can use algorithms such as stochastic gradient descent to train the RBM. However, another problem remains: without access to the partition function $Z$, we cannot compute the expectation with respect to $p(\mathbf{v}, \mathbf{h})$ efficiently. 

To compute the likelihood gradient, Eq.~\eqref{eq:gradient_full}, one can use a sampling algorithm to approximate the expectation with respect to $p(\mathbf{v}, \mathbf{h})$. The most widely adopted algorithm, contrastive divergence (CD)~\cite{goodfellow2016deep}, starts from a sample $\mathbf{v}^0$ from the dataset and constructs a Markov chain of samples,
\begin{equation}
    \mathbf{v}^0\to\mathbf{h}^0\to\mathbf{v}^1\to\mathbf{h}^1\to\cdots\to\mathbf{v}^k, \label{eq:chain_supp}
\end{equation}
using the conditional distributions $p(\mathbf{h}|\mathbf{v})$ and $p(\mathbf{v}|\mathbf{h})$. When $k\to\infty$, the distribution of $\mathbf{v}^k$ converges to $p(\mathbf{v})$, and we can approximate the second term in Eq.~\eqref{eq:gradient_full} with an expectation over a batch of sampled $\mathbf{v}^k$. 

In practice, $k$ can be chosen to be very small, and the performance of CD-$k$ is reasonable even with $k=1$. In this case, the samples are biased from the actual RBM distribution, and we are actually minimizing the difference between two KL-divergences \cite{hinton2002training} (hence the name contrastive divergence),
\begin{equation}
    \mathrm{KL}(q|p) - \mathrm{KL}(p_k|p)
\end{equation}
where $p_k$ is the distribution of the visible nodes after $k$ steps of the Markov chain. As $k\to\infty$, $p_k\to p$, and the approximation becomes exact. 

However, as we will show next, in some cases, the Markov chain does not reach stationarity even with very large $k$. When this happens, CD practically fails as a useful sampler, and mode-assisted training would be necessary to train the RBM successfully~\cite{manukian2020mode}. 

\section{Distance measure on the RBM}
\label{sec:distance}
In the main text, we mentioned that CD only utilizes {\it local}, short-range information, while mode-assisted training can incorporate {\it global}, long-range information into the training procedure. Here we define precisely what kind of distance we refer to.

Recall that CD utilizes a Markov chain, Eq.~\eqref{eq:chain_supp}, to sample the state space. At the $i$-th step $\mathbf{v}^i\to\mathbf{h}^i\to\mathbf{v}^{i+1}$, the achievable states $\mathbf{v}^{i+1}$ might be limited. In this case, the achievable states $\{\mathbf{v}^{i+1}\}$ become the ``neighborhood'' of $\mathbf{v}^i$. With respect to the RBM, we define the distance between two configurations, $\mathbf{v}^i$ and $\mathbf{v}^{j}$, as

\begin{equation}
    d(\mathbf{v}^i\to\mathbf{v}^{j})=-\log\left(\sum_{\mathbf{h}}p(\mathbf{v}^{j}|\mathbf{h})p(\mathbf{h}|\mathbf{v}^{i})\right), \label{eq:distance}
\end{equation}
which is the negative log transition probability from $\mathbf{v}^i$ to $\mathbf{v}^j$. With this definition, we can verify that $d(\mathbf{v}^i\to\mathbf{v}^{j})+d(\mathbf{v}^j\to\mathbf{v}^{k})$ leads to the transition probability $p(\mathbf{v}^i\to\mathbf{v}^j\to\mathbf{v}^k)$. 

Note that the distance defined in Eq.~\eqref{eq:distance} is not symmetric and doesn't necessarily satisfy the triangle inequality. Rather, if we view each basis state in the Hilbert space as a vertex on a graph, Eq.~\eqref{eq:distance} will act as the directed graph distance between two vertices. In this way, we can define the concept of locality on the graph. 

\begin{figure}[htbp]
    \centering
    \includegraphics[width = 0.48\textwidth]{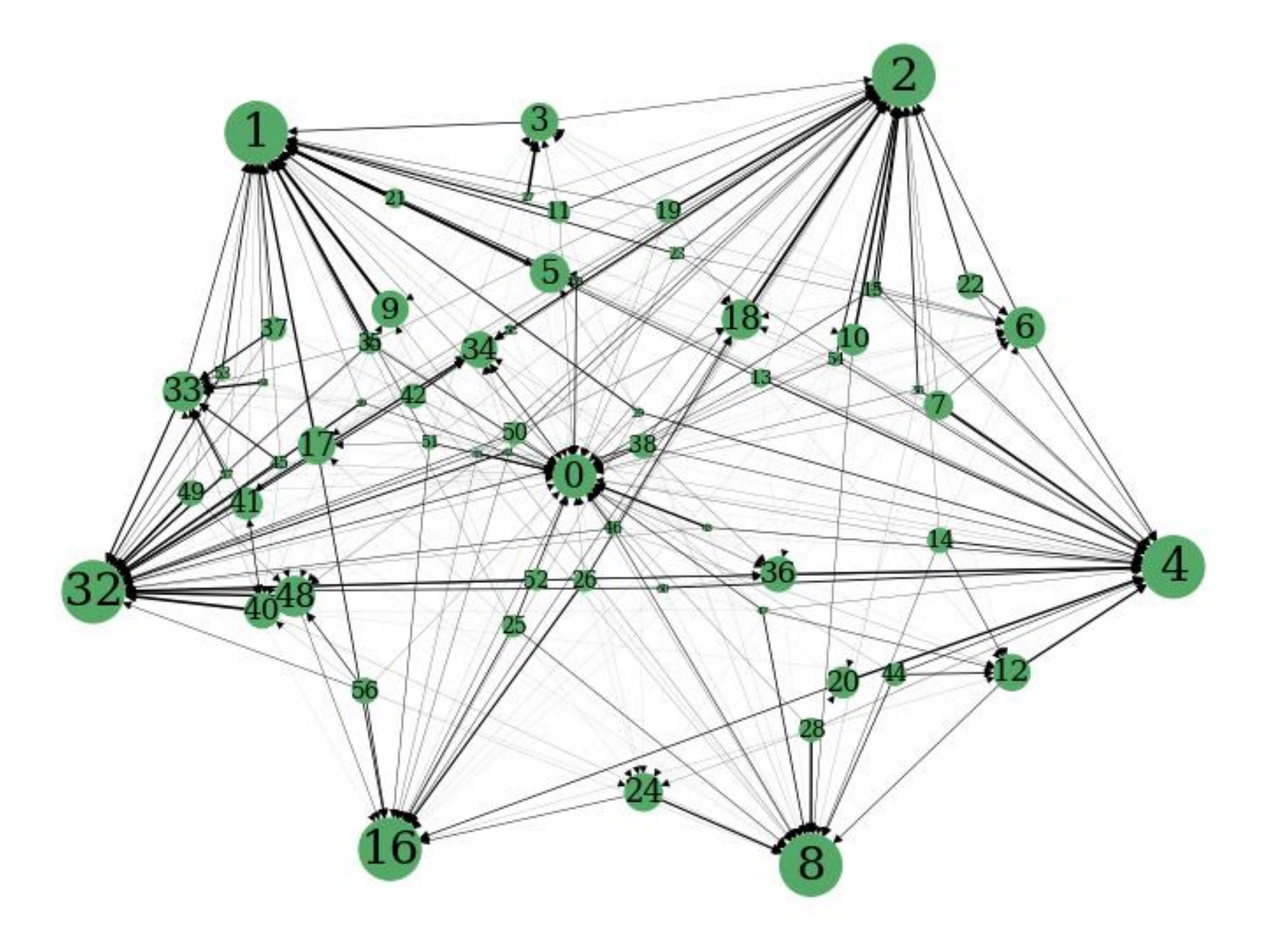}
    \caption{Visualization of the graph structure of an RBM, trained on the 6-qubit W-state. Sizes of vertices represent probabilities in the RBM distribution (not proportional), and width of edges are proportional to the transition probabilities in the Markov chain during sampling. With this graphical representation, we can view CD as random walk on the graph. }
    \label{fig:distance}
\end{figure}

In Fig.~\ref{fig:distance}, we plot the weighted directed graph defined above using the NetworkX python package \cite{SciPyProceedings_11}, with the reference RBM trained on a 6-qubit W-state. Each vertex is numbered using the decimal representation of its corresponding binary basis vector, with larger vertices representing larger probabilities in the RBM distribution, and widths of the edges proportional to the transition probabilities. Edges with transition probabilities less than $0.01$ are omitted. 

The layout of the vertices are computed using the Fruchterman-Reingold force-directed algorithm \cite{fruchterman1991graph}, which treats the vertices as repelling objects and edges as springs holding them close. At equilibrium, the spatial proximity of vertices would more or less characterize the distance between basis vectors in the Hilbert space. 

Recall that $|000001\rangle$ through $|100000\rangle$ are the six most important bases in the W-state. In Fig.~\ref{fig:distance}, the corresponding nodes $1, 2, 4, 8, 16$ and $32$ form a hexagon, enclosing all other nodes in it. None of them have an edge connecting each other \textemdash in fact, none of them even have an outgoing edge at all! According to our defined distance measure Eq.~\eqref{eq:distance}, they are indeed far apart from each other. 

Now, we can view CD-$k$ as random walk on the graph for $k$ steps. Then, it is immediately obvious that, CD is a {\it local sampler}, in the sense that it can only access a small neighborhood of the starting vertex. What is worse, in Fig.~\ref{fig:distance}, the six out-most vertices act as sinks for the random walker: once it gets in one of them, escaping it is almost impossible. In this case, ergodicity is practically lost, and CD fails as a useful sampler.

\section{A further look into CD-\it{k}}
\label{sec:slow_mix}

We further demonstrate via a sampling experiment that, access to global minima cannot be cured with a local sampler, like CD, by simply increasing the length of the Markov chain. 

\begin{figure*}[htbp]
	\centering
	\includegraphics[width = 0.9\textwidth]{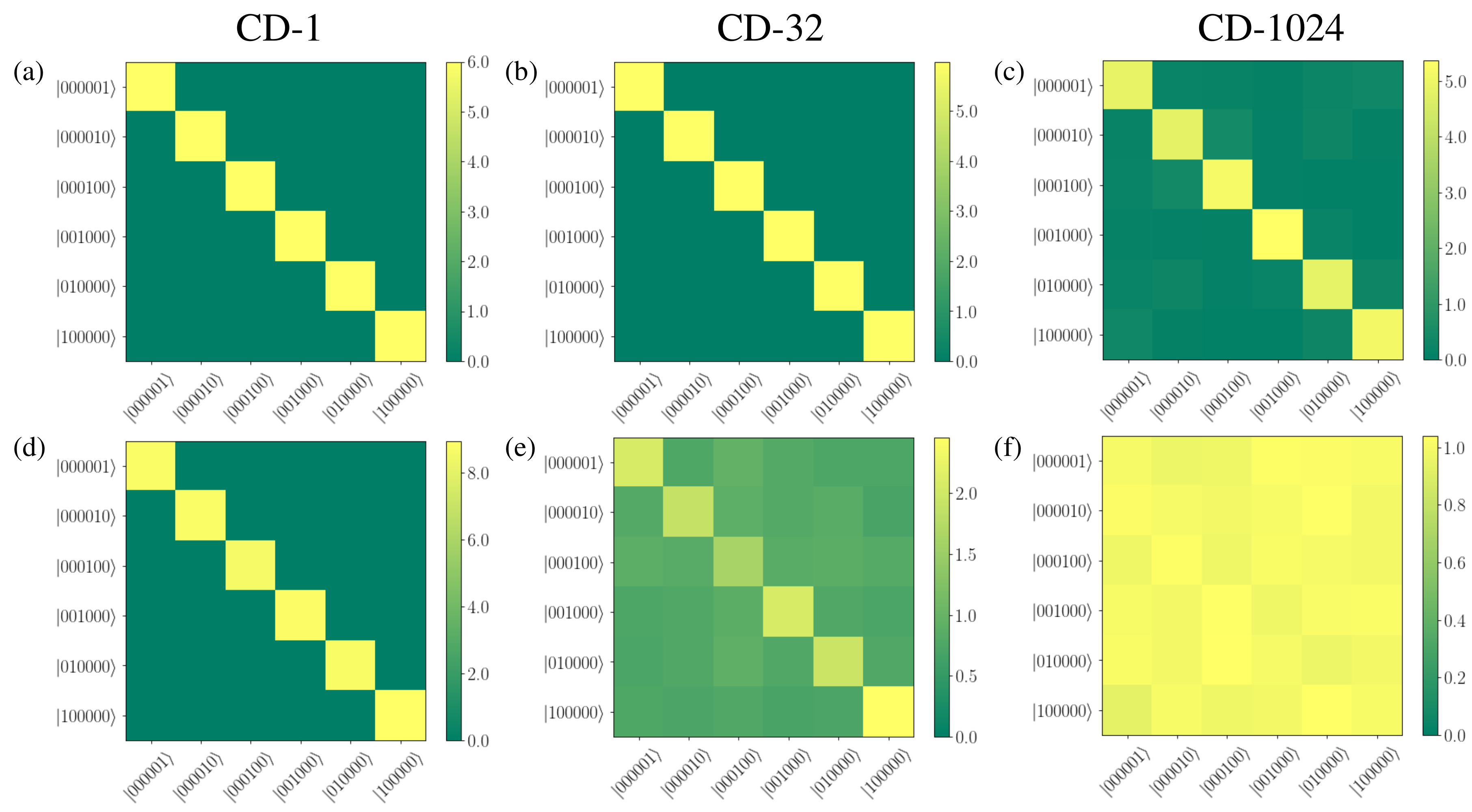}
	\caption{Normalized transition probability when sampling a trained RBM with CD-k. Location $(i, j)$ represents the normalized transition probability, $p_{i\to j}/p_j$, of starting the Markov chain from configuration $i$ and ending in configuration $j$. We observe a strong correlation between the initial and final configuration. (a), (b), and (c) The 6-qubit pure W-state. With isolated modes and zero amplitude on all other basis, the energy barriers between different modes is so high that even CD-1024 cannot escape them. (d), (e), and (f) The depolarized 6-qubit W-state with $p=0.4$. With a background noise, CD can properly escape from each local minimum and explore other parts of the phase space. }
	\label{fig:transition_prob}
\end{figure*}

Again, using the RBM trained on the 6-qubit W-state, we start the sampling chain from one of the modes, and perform CD-$k$ sampling for $10^4$ times. The normalized transition probability is plotted in Fig.~\ref{fig:transition_prob}: the $(i, j)$-th location in each plot represents $p_{i\to j} / p_j$, the probability of starting from the $i$-th state and ending in the $j$-th state, divided by the probability of the $j$-th state in the original distribution. In the ideal case where the Markov chain has sufficiently mixed, the sampled distribution should converge to the exact distribution and be independent of the starting point, resulting in $p_{i\to j} / p_j = 1$.

However, as we clearly see in Fig.~\ref{fig:transition_prob} this is not the case. 
Figures~\ref{fig:transition_prob}(a), (b) and (c) show the results of CD-$k$, with $k=1, 32, 1024$, respectively. Even with CD-1024, most sampling chains are still stuck at their starting point. Practically, CD fails as a sampler in this case, as it cannot properly explore the phase space: when falling into a mode, it cannot easily escape from it. 

\begin{figure}[htbp]
	\centering
	\includegraphics[width = 0.48\textwidth]{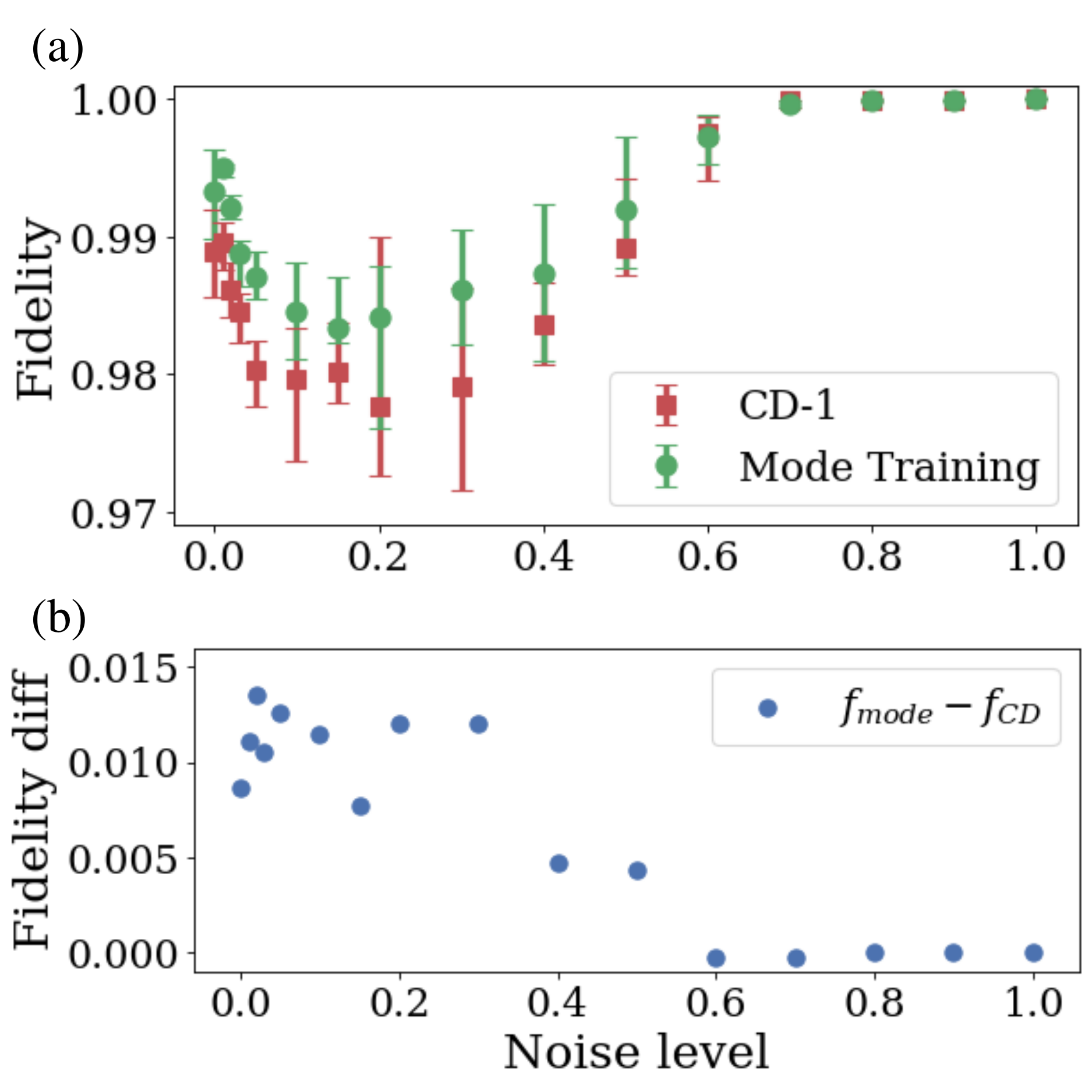}
	\caption{(a) Reconstruction of the depolarized W-state $\rho_W = (1-p)|\Psi_W\rangle\langle\Psi_W| + p \mathds{1} / 2^N$. Data points are medians over 20 runs, and error bars represent corresponding maximum and minimum values. Reconstruction is most difficult at a moderate noise level $p\sim 0.15$. (b) Fidelity difference between mode-assisted training and CD-1. As sampling gets easier with the introduction of the background noise, the advantage of mode-assisted training gradually disappears.  }
	\label{fig:W_noise}
\end{figure}

We can then naturally ask: for what type of distributions does CD work well? To this end, we consider the depolarized W-state, whose density matrix is given by:

\begin{equation}
	\hat{\rho}_W = (1-p)|\Psi_W\rangle\langle\Psi_W| + p \mathds{1} / 2^N
	\label{eq:depolarized_W}
\end{equation}
where $p$ controls the noise level, and $N$ is the number of qubits. To get comparable results to the pure W-state $|\Psi_W\rangle$, we synthesize a similar dataset by perform 2-outcome POVMs described by the measurement operators $M_i=\{|0\rangle\langle 0|, |1\rangle\langle 1|\}$. To be specific, in each measurement, the full measurement operator is the tensor product $M=M_1\otimes M_2\otimes\cdots\otimes M_N$, and the outcome of the measurement can be written as a bitstring $v_1v_2\cdots v_N$. In the noiseless limit, this is the projective measurement in the computational basis, and the outcome bitstrings precisely correspond to the basis vectors $|v_1v_2\cdots v_N\rangle$. With the depolarization noise, the second term in Eq.~\eqref{eq:depolarized_W} acts as a  uniform background noise to the distribution of the measurement outcome. 
Note, however, that this measurement is not informationally complete.

In Figs.~\ref{fig:transition_prob} (d), (e) and (f), we show the same sampling experiment, with $p=0.4$. Thanks to the background noise which lowers the energy barrier between different modes, it now becomes possible for local moves to jump out of the local minima, and the CD chains can properly converge with increasing chain length $k$. 

Fig.~\ref{fig:W_noise} shows this effect from another perspective, where we plot the final fidelity after training versus the noise level $p$, on the depolarized 10-qubit W state. Unlike some works claiming that the noiseless limit is the hardest to learn \cite{carrasquilla2019reconstructing}, here we find that the difficulty peaks near $p\sim 0.15$. We suspect this is due to the fact that the model has to learn the exact amplitude of a small but nonzero background noise, which is more difficult than simply setting the background to zero.

Irrespective, since the background noise greatly eases the burden of sampling, the advantage of mode-assisted training gradually disappears as $p$ increases, until both methods converge to $f=1$ as $p\to 1$. Such strongly noisy states are easier for local samplers, as local moves would be sufficient to explore the entire phase space, and this difficulty is tunable as we change the parameter $p$.  
This further reinforces the notion that approaches providing non-local information to the training (as the one we have discussed here) 
are very important for quantum states with strongly non-local features.

\section{Advanced samplers}
\label{sec:PCD_PT}
To overcome the weaknesses of CD, many advanced sampling algorithms have been proposed, aiming at alleviating the slow mixing problem of the Markov chain. In this section, we examine two notable examples, persistent contrastive divergence (PCD) \cite{tieleman2008training} and parallel tempering (PT) \cite{earl2005parallel, desjardins2010parallel, cho2010parallel}.  

The quality of the samples drawn by CD depends strongly on the length of the Markov chain. Generally, CD-$k$ with a large $k$ performs better than CD-1, at the expense of much longer running time. 

PCD builds on the idea that, instead of starting a new Markov chain for sampling at every training step, one can maintain one Markov chain throughout the entire training process. With a small learning rate, the RBM distribution only changes slightly at every training step. Therefore, if the Markov chain is sufficiently mixed at the previous training step, it will be close enough to equilibrium at the next training step. By maintaining one Markov chain throughout training, one is essentially using CD-$k$, with $k$ very large. 

PT is also known as replica exchange MCMC sampling, which maintains many copies of the system at different temperatures, and exchange of configurations at different temperatures is allowed according to some acceptance criteria. Since higher temperatures allow the system to explore the high energy configurations more efficiently, the resulting algorithm is less prone to getting stuck in local minima. 

For training RBMs, PT maintains $N$ copies of the same RBM, with the distributions 
\begin{equation}
    p_i(\mathbf{v})=\frac{e^{-\beta_i E(\mathbf{v})}}{Z}
    \label{eq:p_PT}
\end{equation}
for a set of gradually increasing temperatures $\{T_1, T_2, \cdots, T_N\}$, with $\beta_i=1/T_i$ denoting the inverse temperature. $T_1=1$ corresponds to the original RBM distribution, and $T_N$ is usually chosen to be a very large number (e.g., $T_N=100$) to ensure a proper exploration of the entire phase space. When running the Markov chain, each copy of the system evolves on its own using Gibbs sampling, and an additional cross-temperature state swap move is introduced. At each sampling step, two neighboring configurations $\mathbf{v}_i, \mathbf{v}_{i+1}$ are exchanged with probability
\begin{equation}
    r=\frac{p_i(\mathbf{v}_{i+1})p_{i+1}(\mathbf{v}_i)}{p_i(\mathbf{v}_{i})p_{i+1}(\mathbf{v}_{i+1})}. \label{eq:PT_accept}
\end{equation}

Using Eq.~\eqref{eq:p_PT}, Eq.~\eqref{eq:PT_accept} becomes
\begin{equation}
    r=\exp\big((\beta_i-\beta_{i+1})(E(\mathbf{v}_i)-E(\mathbf{v}_{i+1})\big).
\end{equation}

\begin{figure}[htbp]
	\centering
	\includegraphics[width = 0.45\textwidth]{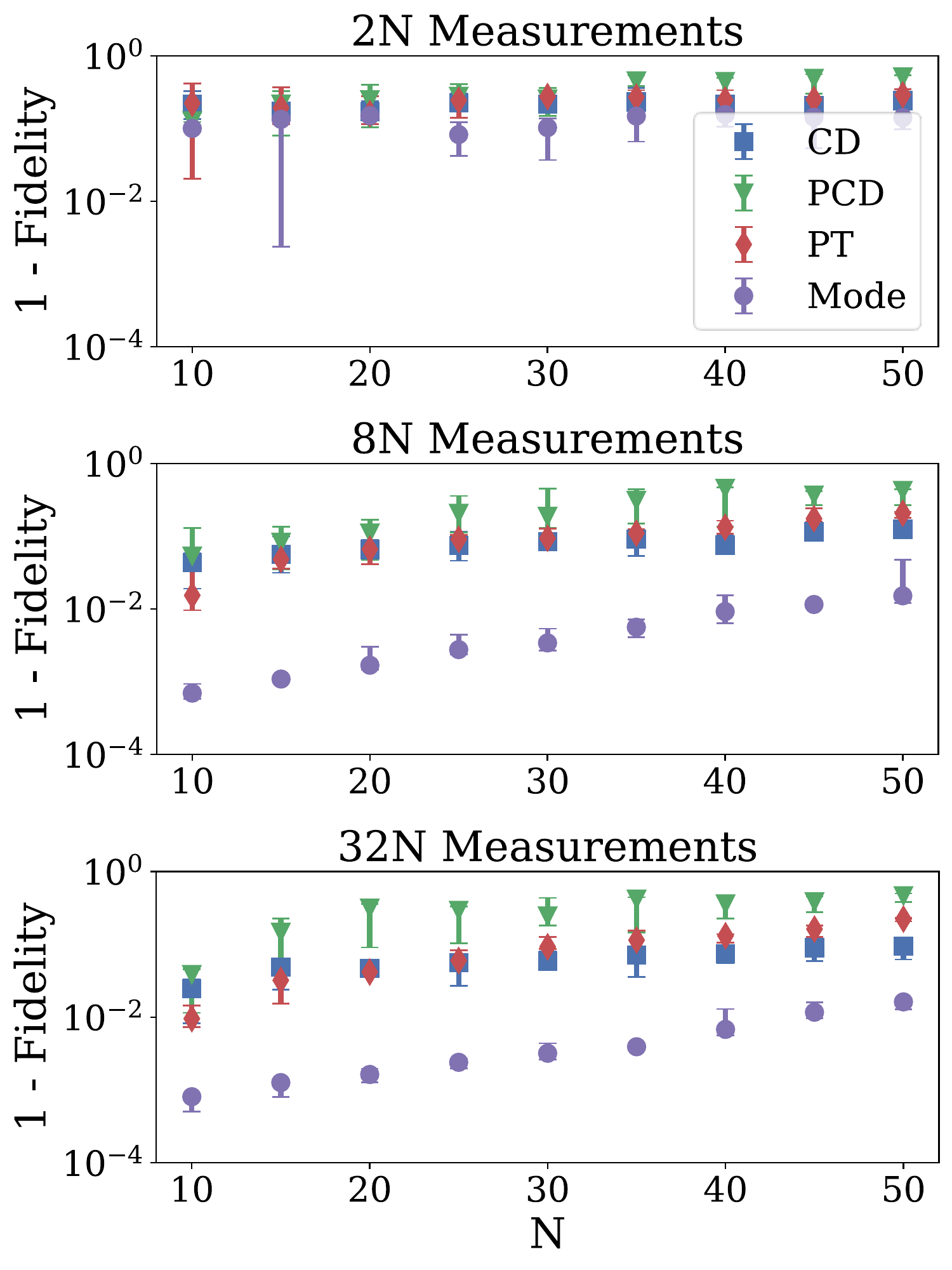}
	\caption{Comparison of 4 different training methods on the W-state. Data points are medians of 5 runs, and error bars represent the maximum and minimum values. Mode-assisted training consistently outperforms other methods.}
	\label{fig:samplers}
\end{figure}

While PCD and PT have already seen some success at training RBMs \cite{tieleman2008training, earl2005parallel, desjardins2010parallel, cho2010parallel}, here, we show that they are not as effective in our case. Fig.~\ref{fig:samplers} compares the performance of CD, PCD, PT and mode-assisted training, on the $N$-qubit W-state. With enough measurements, mode-assisted training consistently outperforms all other methods by at least one order of magnitude. PCD has the worst performance, and PT only outperforms CD on smaller systems.

PCD and PT are designed to improve on CD, but why are we seeing worse performance here? To understand this behavior, let us again focus on Fig.~\ref{fig:distance}. While PCD and PT are designed to alleviate the slow mixing problem, their underlying proposal and acceptance steps are the same as CD. Therefore, they are still random walkers on Fig.~\ref{fig:distance}, except with a new set of random walk rules, and they suffer from the same problem as CD: the random walker will get stuck on one mode configuration, unable to escape, and ergodicity is lost. Importantly, PCD and PT start from random initial conditions, and the distribution of the ending mode state they get stuck in is likely biased from the RBM distribution. While CD also gets stuck, it is initialized with configurations from the dataset, and the ending distribution will be much closer to the actual RBM distribution. 

Again, we see the superiority of mode-assisted training, and the usefulness of global information. While PCD and PT may offer improvements compared to CD in certain cases, they are still local samplers, in the sense that they only utilize local information, performing a random walk on a small region of the graph. One potential improvement for them could be explicitly providing global information at the proposal-acceptance step, like mode-hopping moves \cite{sminchisescu2003mode}. But again, doing so requires prior knowledge to the target distribution, while mode-assisted training achieves this automatically.

\section{Further numerical experiments on entangled quantum systems}
\label{sec:numerical_results}

In this section, we further test the capabilities of mode-assisted training on two more highly-entangled quantum systems: the transverse-field frustrated Ising model (TFFIM) on the triangular lattice \cite{moessner2001ising, wang2017caution}, and the toric code model \cite{kitaev2003fault}. As a proof-of-concept demonstration, we focus on small systems where exact results are available using exact diagonalization, and compare the performance between mode-assisted training and CD-1. 

Fig.~\ref{fig:triangular} is an illustration of the TFFIM on the triangular lattice. The Hamiltonian reads: 

\begin{figure}
	\centering
	\includegraphics[width = 0.45\textwidth]{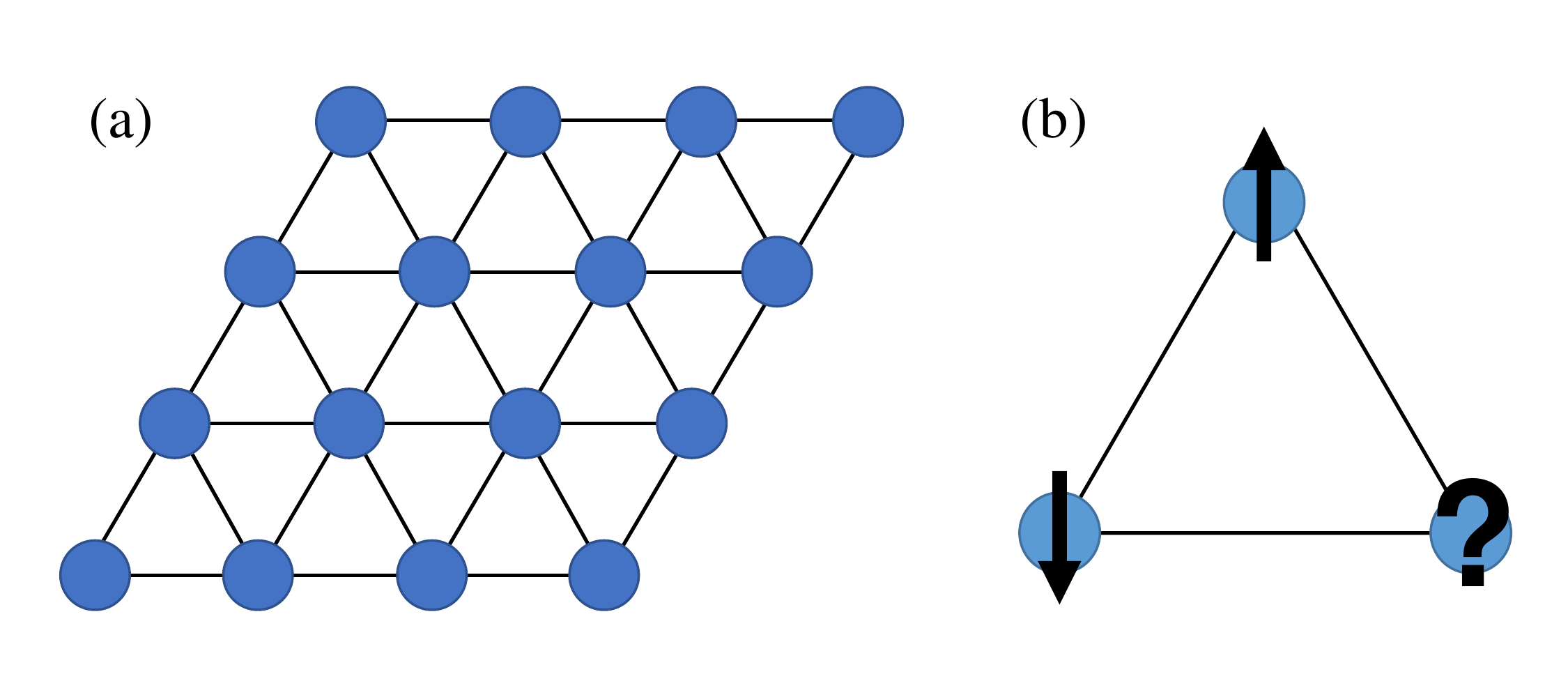}
	\caption{(a) An illustration of the triangular lattice. (b) A frustrated loop. With antiferromagnetic interactions, two neighboring spins tend to anti-align with each other, leaving the direction of the third spin undetermined. }
	\label{fig:triangular}
\end{figure}

\begin{equation}
    H=J\sum_{\langle i, j\rangle} \sigma_i^z \sigma_j^z - h\sum_i \sigma_i^x,
\end{equation}
where $J$ is the nearest-neighbor antiferromagnetic Ising coupling and $h$ is the transverse field. In this demonstration, we choose $h=J=1$. 

Without the transverse field, frustration would lead to a highly degenerate ground state \cite{kirkpatrick1977frustration}. Together with the transverse field, we arrive at a ground state wave function multi-modal in the $\sigma^z$ basis. 

\begin{figure}[htbp]
	\centering
	\includegraphics[width = 0.45\textwidth]{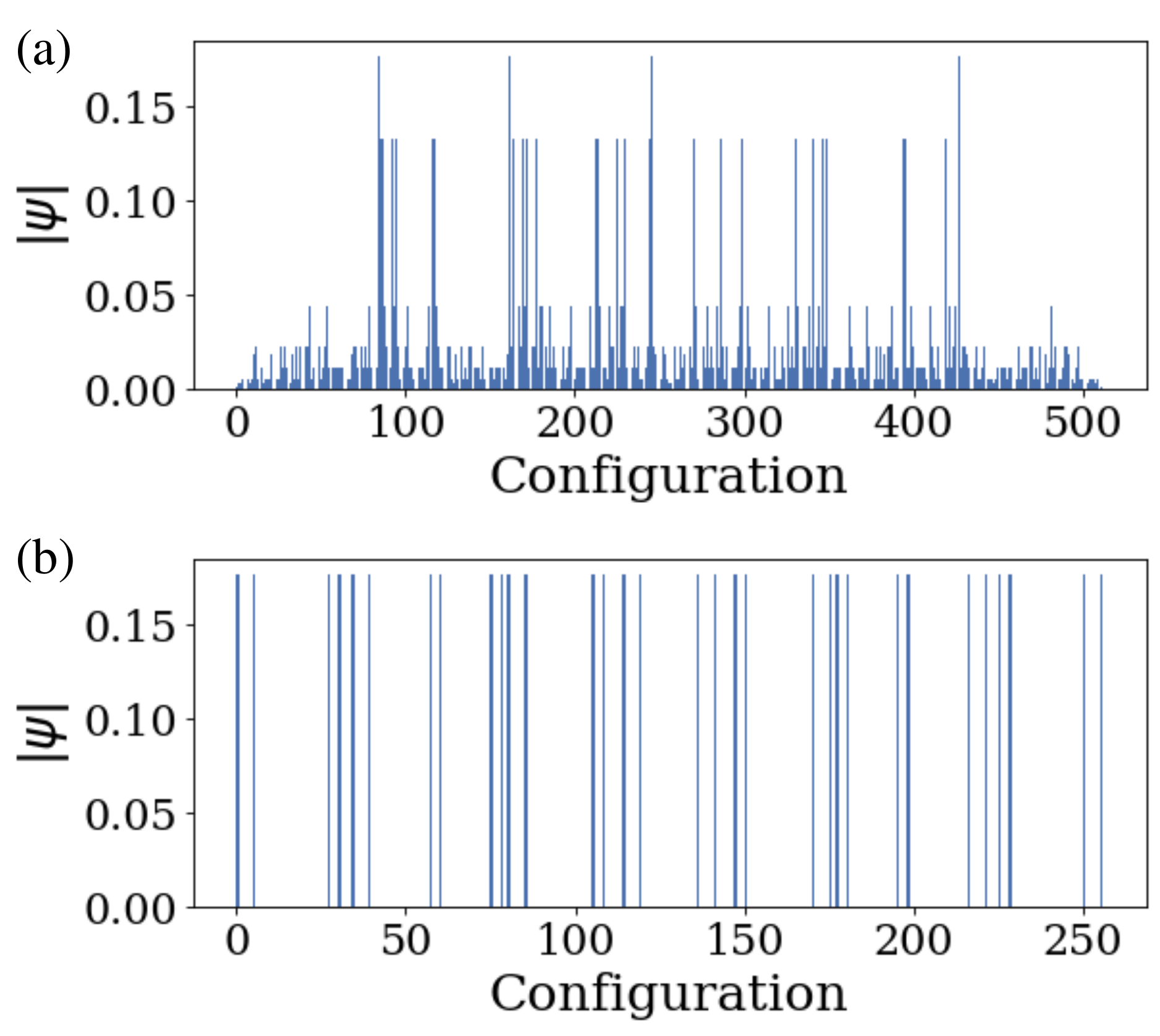}
	\caption{(a) The ground state wave function of the $3\times 3$ TFFIM. (b) One ground state of the $2\times 2$ toric code model. While both distributions are multi-modal, the first distribution has a structured background, making jumps between modes easier. On the contrary, the second distribution has isolated modes, making it difficult for local samplers to capture the entire distribution. }
	\label{fig:psi}
\end{figure}

In Fig.~\ref{fig:psi} (a), we plot the exact ground state of the $4\times 4$ TFFIM on the triangular lattice. A synthetic dataset is generated by taking $10^4$ projective measurements on this state, and we perform quantum state tomography on it using methods described in the main text. 

\begin{figure}[htbp]
	\centering
	\includegraphics[width = 0.45\textwidth]{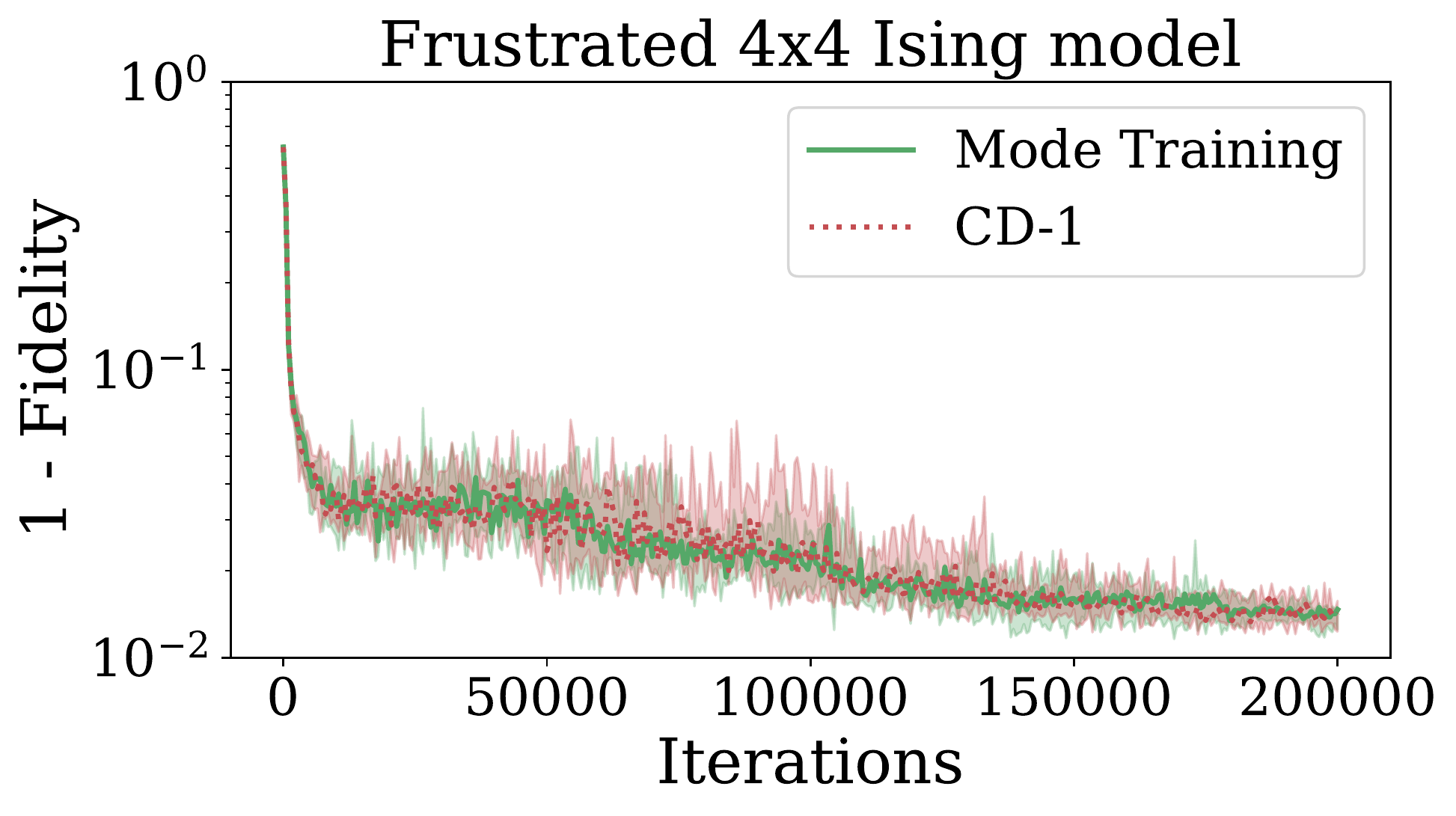}
	\caption{The training curve on the $4\times 4$ TFFIM. Since sampling from this state is easy, mode-assisted training and CD-1 have comparable performance. }
	\label{fig:Ising2D}
\end{figure}

The result is shown in Fig.~\ref{fig:Ising2D}: the performance of mode-assisted training and CD-1 are comparable. This is an example where CD already works very well, and we would not gain much from mode-assisted training. The reason is similar to the frustrated W-state we just showed: the background structure in the distribution in Fig.~\ref{fig:psi} (a) helps the sampler to jump between different modes, and the quality of the samples is already good using CD-1. In such cases, mode-assisted training would be an overkill. 

Another example is the toric code model \cite{kitaev2003fault}. It was already demonstrated that an RBM can exactly and efficiently represent the ground state of the toric code, with analytically computable RBM weights \cite{deng2017machine, jia2019efficient, zhang2018efficient}. However, the story is quite different if we actually train an RBM with data sampled from the toric code state, which is what it would happen in an actual experiment. 

\begin{figure}[H]
	\centering
	\includegraphics[width = 0.45\textwidth]{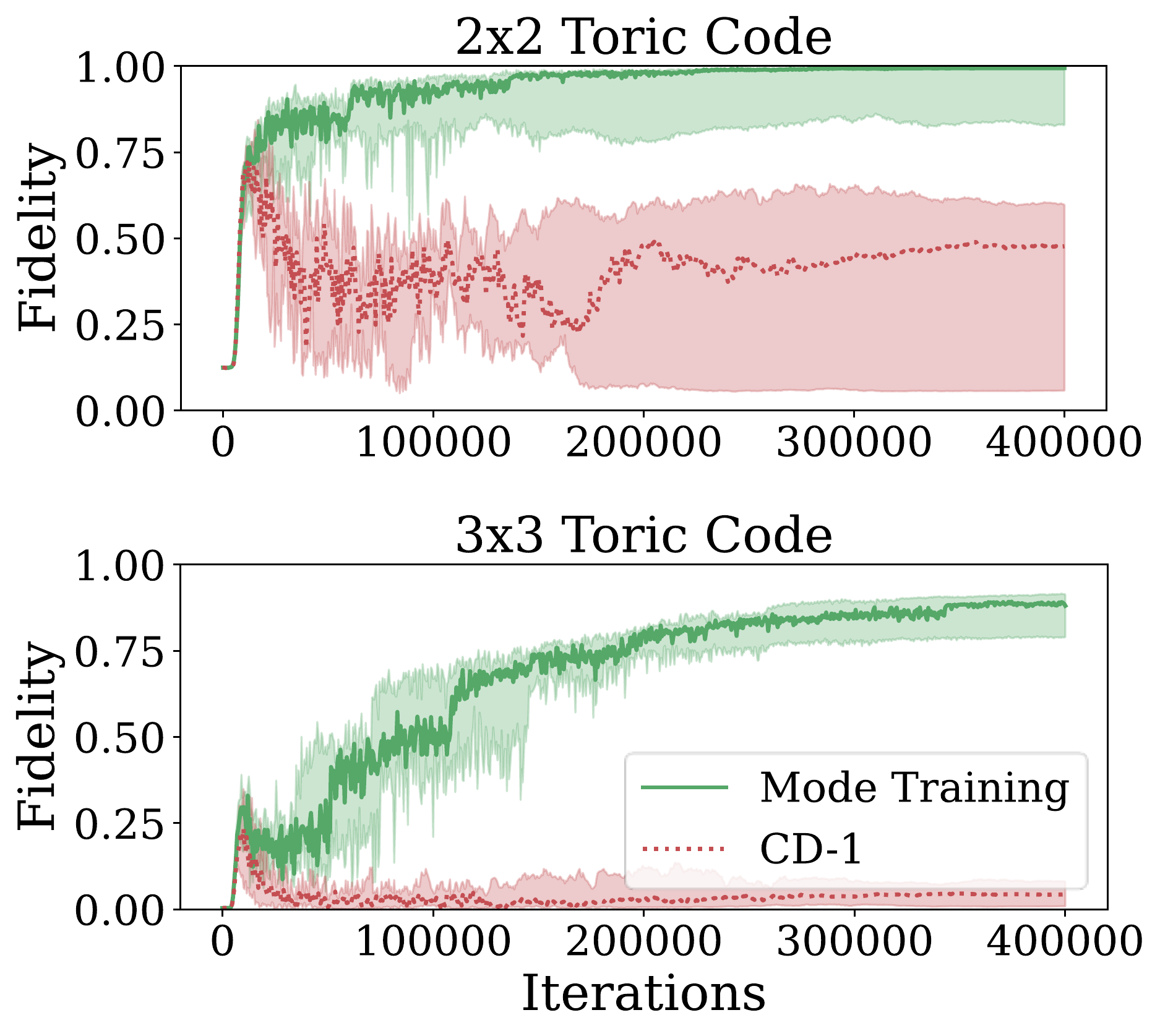}
	\caption{The training curve of the $2\times 2$ and $3\times 3$ toric code state. CD-1 completely fails, while mode-assisted training still performs reasonably well. }
	\label{fig:toric}
\end{figure}

Fig.~\ref{fig:psi} (b) shows one ground state of the $2\times 2$ toric code model: a multi-modal distribution with isolated modes, which, according to our analysis, is difficult for local samplers such as CD. The training curve in Fig.~\ref{fig:toric} confirms this prediction. Even for the $2\times 2$ toric code state with 8 qubits, CD-1 performs rather poorly, while mode-assisted training can achieve near perfect fidelity. And in the $3\times 3$ case, CD-1 completely fails with final fidelity less than 0.1, but mode-assisted training can still reasonably reconstruct this state with fidelity near 0.9. 

We can now clearly understand when the mode training is particularly advantageous. If a local sampler like CD performs well when training the RBM, then mode-assisted training would not be very useful. But for multi-modal states with strongly non-local features, mode-assisted training can offer significant advantages over traditional methods.

\end{document}